\documentstyle[prl,aps]{revtex}
\begin{document}
\draft \title{Orbital dependent superconductivity in Sr$_2$RuO$_4$}
\author{D.F. Agterberg, T.M. Rice, and M. Sigrist}
\address{Theoretische Physik, Eidgenossiche Technische Hochschule-
H\"{o}nggerberg, 8093 Z\"{u}rich, Switzerland}
\date{\today}
\maketitle
\begin{abstract}
We show that for  
superconducting
Sr$_2$RuO$_4$ any unconventional pairing in   
the part of the Fermi surface with Ru $4d_{xy}$ orbital
character
is weakly coupled to that with
Ru $\{4d_{xz},4d_{yz}\}$ orbital character. This naturally
gives rise to two disparate energy scales in the 
superconducting state which leads to novel low temperature properties
in a variety of thermodynamic and transport properties and which
would also account
for the large residual density of states seen in specific heat and NQR
measurements.
\end{abstract}


Sr$_2$RuO$_4$ provides the 
first example
of a layered perovskite material that exhibits superconductivity 
without the presence of copper \cite{mae94}.
Even though there is a close structural similarity with the high $T_c$ 
materials, the electronic properties are very different. 
While it 
is clear that electron correlation effects are important in Sr$_2$RuO$_4$,
the normal state near the superconducting transition is well described by 
a quasi-2D Landau Fermi liquid ({\it e.g.} the resistivity in all directions
follows a $T^2$ behavior for $T\lesssim 50$ K and the 
resistivity along and perpendicular to the $c$ axis differ by a factor
of 850 \cite{mae94}). 
Quantum oscillations show
three Fermi surface sheets with a 2D topology 
that agrees
well with band structure calculations \cite{mac296}.
It has been pointed out \cite{ric95} that the mass enhancement is similar to
that of $^3$He and that there is a metallic
ferromagnetic phase in SrRuO$_3$ \cite{gib74} (the 3D analogue
of Sr$_2$RuO$_4$). These observations indicate
that an odd-parity ($l=1$) superconducting state is likely \cite{ric95}.
 This is consistent with the lack of a 
Hebel-Slichter peak in $1/T_1$ in NQR measurements \cite{ish97}.
A weak coupling analysis of the odd-parity state implies the gap 
should be of constant magnitude \cite{ric95}.
It is therefore surprising that specific heat 
\cite{mae97} and NQR measurements \cite{ish97} reveal that 
approximately 0.6 of the normal density of states 
remain in the superconducting phase in clean samples (those
in which quantum oscillations were observed).
As a consequence it has been proposed that 
an exotic non-unitary superconducting state similar
to the $^3$He $A_1$ phase is stabilized\cite{sig96,mac96}. 
In this scheme, the normal
state quasiparticle energy  spontaneously splits into two branches 
(one for spin up and 
one for spin down) upon entering
the superconducting state. 
One of these branches is gapped while 
the other is not, leading to a residual density of states that
is half the normal density of states.

Here we propose
an alternative explanation for the large residual
density of states. The electronic properties near the Fermi surface 
of Sr$_2$RuO$_4$ are determined by Wannier functions with Ru
$d_{xy},d_{xz}$, and $d_{yz}$ orbital character
\cite{ogu95,sin95}. We show that the quasi-2D
nature of the electronic dispersion implies that the 
bands are derived from either the $xy$ or the
$\{{xz},{yz}\}$ Wannier functions and that 
the pair scattering amplitude between these two classes 
of bands will be significantly 
smaller 
than the intraclass pair scattering amplitude for unconventional 
superconducting order parameters. It can
therefore be expected that the gap on bands from one class is 
substantially smaller 
than that on bands from the other class. The presence of 
essentially gapless excitations 
for temperatures
greater than the smaller gap will appear as a residual density 
of states. Also, the two classes may 
favor different
superconducting symmetries in which case a second superconducting 
transition will 
appear at low temperatures.
    
\begin{figure}[ht]
\vspace{7.0 cm} \relax \noindent \relax
\includegraphics{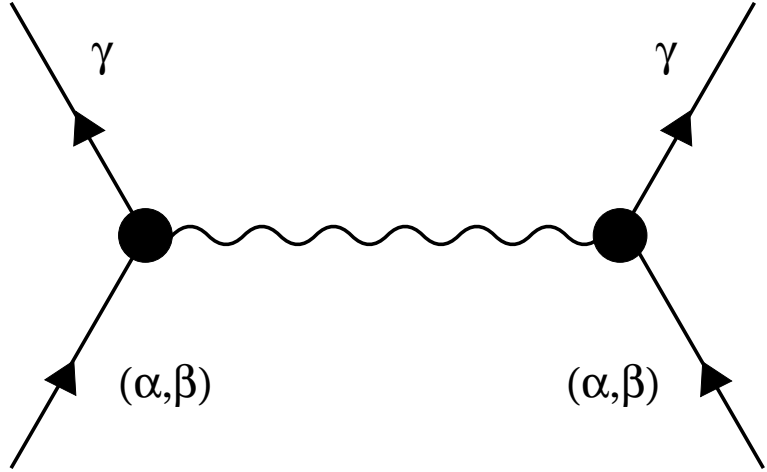}
\caption{{\tenrm \baselineskip=11pt The vertex leading to the pair
scattering amplitude between the $\gamma$ sheet and the other 
two sheets of the Fermi
surface. The effective interaction for any unconventional 
gap symmetry due to this
vertex is small in relation to intrasheet interactions.}}
\label{fig1}
\end{figure}

Band structure calculations 
\cite{ogu95,sin95} 
give the correct shape of the Fermi surface, but 
predict an effective mass that is a factor of 4 smaller than that observed;
indicating that strong coupling effects are important \cite{mac296}. 
These
calculations reveal that the 
density of states
near the Fermi surface are due mainly to the four Ru $4d$ electrons 
in 
the $t_{2g}$ orbitals. 
There is a strong hybridization of these orbitals with the O
$2p$ orbitals giving rise to antibonding $\pi^*$ bands. The resulting
bands have three quasi-2D Fermi surface sheets labeled $\alpha,\beta,$
and $\gamma$ (see Ref. \cite{mac296}). The highly anisotropic nature
of the Fermi liquid and the superconducting states suggests that the
superconductivity essentially arises from intraplanar interactions, so
we consider a single RuO$_4$ plane.
The Hamiltonian describing the band structure of a plane is 
\begin{equation}
H =\sum_{\nu,\nu^{\prime},i,j, s} 
t_{\nu,\nu^{\prime}}({\bf R}_i-{\bf R}_j)
c^{\dagger}_{\nu,i,s}c_{\nu^{\prime},j,s}
\label{Eq1}
\end{equation}
where $c_{\nu,i,s}$ destroys an electron with spin $s$ 
 in the Wannier function centered at ${\bf R}_i$ that transforms
as the Ru $\nu$ orbital ($\nu=\{xy,xz,yz\}$). 
Due to the $\sigma_z$ reflection symmetry about the center of the RuO$_4$ 
plane  
$t_{xy,xz}({\bf R})=t_{xy,yz}({\bf R})=0$.  
This implies that the 
$\gamma$ sheet
of the Fermi surface
can be attributed solely to the ${xy}$ Wannier functions while the $\alpha$ and
$\beta$ sheets are due to a hybridization of the 
$\{{xz},{yz}\}$ Wannier functions. 
An effective Hamiltonian to describe the superconductivity is
\onecolumn
\begin{equation}
H= \sum_{l, {\bf k}, s} \epsilon_{l}({\bf k})
a^{\dagger}_{l,{\bf k},s}a_{l,{\bf k},s}
+
\sum_{l,l^{\prime},{\bf k}, {\bf k^{\prime}},s,s^{\prime}} 
[V_{l,l^{\prime}}({\bf k},{\bf k}^{\prime})
a^{\dagger}_{l,{\bf k},s} 
a^{\dagger}_{l,-{\bf k},s^{\prime}}
a_{l^{\prime},-{\bf k}^{\prime},s^{\prime}}
a_{l^{\prime},
{\bf k}^{\prime},s} + h.c.]
\end{equation}
where $a_{l,{\bf k},s}$ corresponds to the eigenoperators of Eq.~(\ref{Eq1})
and 
\begin{equation}
V_{l,l^{\prime}}({\bf k},{\bf k}^{\prime})=\int d^3r d^3r^{\prime}
\sum_{j,j^{\prime},n,n^{\prime}}  
e^{i{\bf k}\cdot({\bf R}_j-{\bf R}_{j^{\prime}})}
\phi^*_l({\bf r}-{\bf R}_j)\phi^*_l({\bf r}^{\prime}
-{\bf R}_{j^\prime}) 
U({\bf r},{\bf r}^{\prime})e^{i{\bf k}^{\prime}\cdot
({\bf R}_n-{\bf R}_{n^{\prime}})}
\phi_{l^{\prime}}({\bf r}^{\prime}-{\bf R}_n)\phi_{l^{\prime}}
({\bf r}-{\bf R}_{n^{\prime}})
\label{eqv1}
\end{equation}
where $U({\bf r},{\bf r}^{\prime})$ is an effective interaction 
and the spatial extent of the Wannier
functions along the $c$ axis restricts the integrations along $z$ and
$z^{\prime}$ to lie near the RuO$_4$ plane.
For the matrix elements  $V_{\gamma,\alpha}$ and
$V_{\gamma,\beta}$ the symmetry of the Wannier
functions under $\sigma_z$ can be exploited to write
\begin{equation}
4U({\bf r}_{\perp},z;{\bf r}_{\perp}^{\prime},z^{\prime})=
2U({\bf r}_{\perp},z;{\bf r}_{\perp}^{\prime},z^{\prime})-
U({\bf r}_{\perp},-z;{\bf r}_{\perp}^{\prime},z^{\prime})-
U({\bf r}_{\perp},z;{\bf r}_{\perp}^{\prime},-z^{\prime}).
\label{eqv}
\end{equation}
The $z$ dependence of the ${xy}$ Wannier functions limits the
integrations along the $z$ direction in the $V_{\gamma,\alpha}$ and 
$V_{\gamma,\beta}$ matrix elements
to a distance on the order of
$l/7$ \cite{exp} where $l$ 
is the distance between two neighboring Ru 
ions. As a consequence,  
the lowest order term in a Taylor series expansion of
Eq.~\ref{eqv} in 
$z/|{\bf r}_{\perp}|$ and $z^{\prime}/|{\bf r}^{\prime}_{\perp}|$ 
will give the largest contribution to  $V_{\gamma,\alpha}$ and 
$V_{\gamma,\beta}$ for all
but the onsite portion (${\bf R}_j={\bf R}_{j^{\prime}}={\bf R}_{m}=
{\bf R}_{m^{\prime}}$) in Eq.~\ref{eqv1}. The lowest non-zero
term is of second order in this 
expansion. Since the on-site contribution is independent of 
${\bf k}$ and ${\bf k}^{\prime}$ it does not contribute to the effective 
coupling
constant for unconventional gap functions. It is therefore expected that 
the pair scattering amplitude 
between the $\gamma$ sheet  and the $\{\alpha,\beta\}$ sheets is 
significantly smaller that the intrasheet pair scattering amplitude 
(see Fig.\ref{fig1}).
Furthermore, since the Wannier functions 
forming the two classes of bands are of different symmetry, 
the intrasheet
pair scattering amplitudes will in general be different. We assume that 
the superconducting state is odd-parity due to the considerations of Ref.\cite{ric95}. Note that the simplest tight binding
approximation to the band structure (in which the Ru  $\{d_{xz},d_{yz}\}$ 
orbitals
overlap only with neighboring O $p-\pi$ orbitals \cite{ogu95}) indicates that 
the
gaps on the $\alpha$ and $\beta$ sheets
are the same magnitude for odd-parity
pairing and we therefore assume that 
the gaps within this class have the same magnitude.

We consider a model in which the
three Fermi surface sheets have cylindrical symmetry and densities of 
states as in Ref. \cite{mac296}. 
We use a weak coupling approach and in accordance with the above
considerations take 
$V_{l,l^{\prime}}({\bf k},{\bf k}^{\prime})=U_{l,l^{\prime}}{\bf k}\cdot
{\bf k}^{\prime}/(\langle k_x^2\rangle_l
\langle k_x^2\rangle_{l^{\prime}})^{1/2}$ where $\langle k_i^2 \rangle_l$
is the average of $k_i^2$ on sheet $l$ and  
\begin{equation}
U=\pmatrix{u_{xy}&u_m&u_m\cr u_m&u & u \cr u_m &u&
u \cr}
\end{equation} 
where the matrix $U$ operates on a basis with components that 
correspond to the Fermi surface sheets $\gamma,\alpha,\beta$ respectively.
Introducing the gap matrix
\begin{equation}
\Delta_{s_1,s_2}(l,{\bf k})=\sum_{{\bf k}^{\prime},l^{\prime}}V_{l,l^{\prime}}({\bf k},{\bf k}^{\prime})
F_{s_1,s_2}(l^{\prime},{\bf k}^{\prime})
\end{equation}
where $F_{s_1,s_2}(l,{\bf k})=\langle a_{l,{\bf k},s_1} a_{l,-{\bf k},s_2} 
\rangle$,
gives rise to a mean field Hamiltonian that is diagonal in the band index.
For an odd-parity interaction the gap can be
expressed as
$\hat{\Delta}(l,{\bf k})=i[{\bf d}_l({\bf k})\cdot 
{\mbox{\boldmath{$ \sigma$}}}
]\sigma_y$ \cite{sig91}.
For unitary states (the case considered here) 
the quasiparticle excitations 
are given by
$E_{l,{\bf k}}=
(\epsilon^2_{l,{\bf k}}+|{\bf d}_l({\bf k})|^2)^{1/2}$
and the gap equation is given by
\begin{equation}
{\bf d}_l({\bf k})=\sum_{{\bf k}^{\prime},l^{\prime}}
\frac{V_{l,l^{\prime}}({\bf k},{\bf k}^{\prime})
{\bf d}_{l^{\prime}}({\bf k}^{\prime})}
{2E_{l^{\prime},{\bf k}^{\prime}}}
\tanh(\beta E_{l^{\prime},{\bf k}^{\prime}}/2).
\end{equation} 
Within weak coupling the transition temperature is
$T_c=1.13 \epsilon_c\exp[-1/\lambda_{max}]$ where 
$\lambda_{max}$ is the
largest eigenvalue of the matrix with components 
$U_{l,l^{\prime}}(N_lN_{l^{\prime}})^{1/2}$ and $N_l$ is
density of states of sheet $l$. It has been assumed 
that the
cut-off frequency $\epsilon_c$ is the same for all three bands. 

The superconducting order parameter is 
${\bf d}_l({\bf k})=\sum_{i,j}c_{l,i,j}k_i \hat{x}_j/\langle 2k_i^2\rangle_l$ 
which has a six fold degeneracy that is broken by spin-orbit coupling.
 The phases stabilized within weak coupling
for the single band version of this model are the planar and the 
axial phases, both are degenerate within the approximations made above 
\cite{sig96,mar87}. Spin-orbit
coupling will prefer one of these two phases and will fix the spin
orientation of this phase to the crystallographic axes, leading 
to the classification in Ref. \cite{ric95}. 
The quasiparticle excitation spectra 
for the possible phases are 
described by a gap of constant magnitude, so many properties will be
correctly described by assuming that any one of
these phases are stabilized. We assume that 
the $A_{1u}$ phase, for which 
${\bf d}_l({\bf k})=c_l(\hat{x}k_x+\hat{y}k_y)/\langle
2 k_x^2\rangle ^{1/2}$, is stabilized. 
The resulting gap equation for the $\{c_l\}$ then has 
the same form as that for isotropic superconductors generalized 
to include the presence 
of three bands \cite{suh59}. 

\begin{figure}[ht]
\vspace{11.0 cm} \relax \noindent \relax
\includegraphics{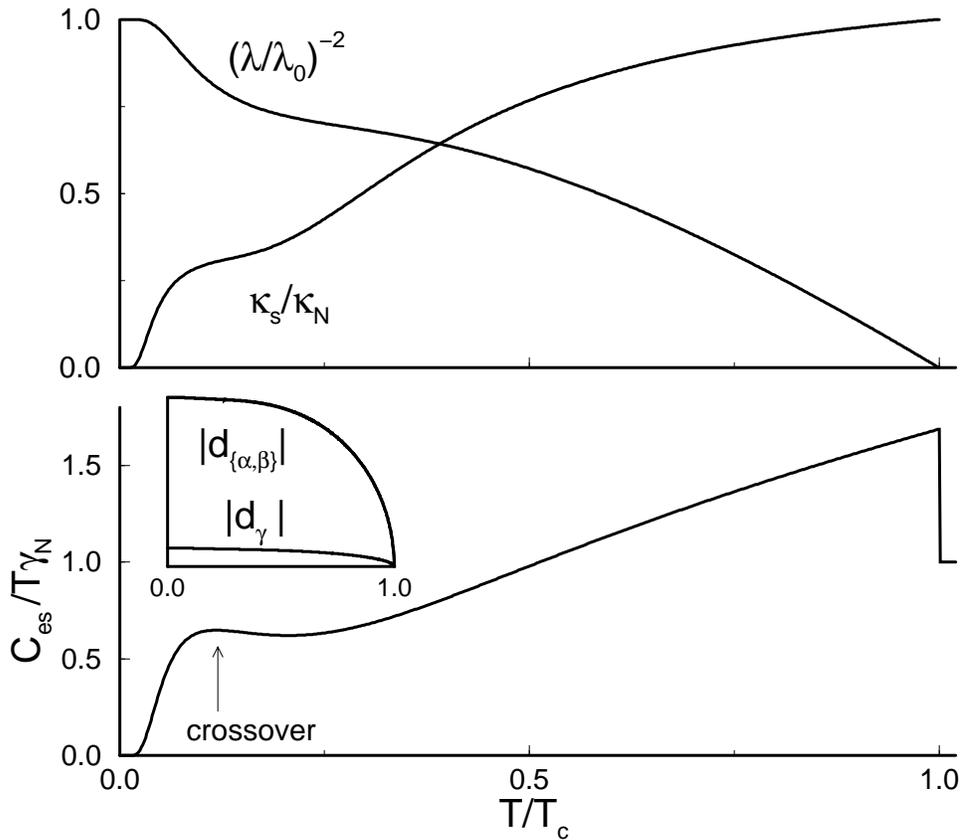}
\caption{{\tenrm \baselineskip=11pt Specific heat, London
penetration depth, and thermal conductivity  as a function
of temperature. The inset shows the magnitude of the 
gaps $\bf{d}_{\gamma}$ and $\bf{d}_{\{\alpha,\gamma\}}$ 
as a function of $T/T_c$. 
}}
\label{fig2}
\end{figure}

The interaction parameters $u_{xy},u_m$, and $u$
remain to be specified. Earlier arguments imply $u_m\ll max(u_{xy},u)$
but the relation between $u_{xy}$ and $u$ remains unknown. 
Hund's Rule ferromagnetic correlations between the Ru 
$d_{xz}$ and $d_{yz}$ orbitals
may give rise to an increased odd-parity interaction 
for the $\{\alpha,\beta\}$ Fermi surface sheets. 
Also, the $\gamma$ sheet is more 
2D than the
$\{\alpha,\beta\}$ sheets, so fluctuations may lead to a 
greater reduction in the $T_c$ for the $\gamma$ than for the 
$\{\alpha,\beta\}$ sheets. These considerations indicate that 
$u>u_{xy}$, so for illustration purposes we
consider this to be the case (though it cannot 
be ruled out that $u_{xy}>u$ without a more detailed microscopic model). 
To show the qualitative behavior of 
this above model we take the density of states as measured in Ref.\cite{mac296}
($N_{\alpha}:N_{\beta}:N_{\gamma}=0.15:0.3:0.55$) and the
following values for the interaction matrix $U$;
$u_{xy}:u_m:u=0.09:0.09:1.0$ with $u N_{\beta}= 0.630$. 
Using for the  specific heat $C_{es}$ 
\begin{equation}
C_{es}=2\sum_{l,{\bf k}} E_{l,{\bf k}} \frac{\partial f(E_{l,{\bf k}})}
{\partial E_{l,{\bf k}}},
\end{equation}
and solving the gap equation yields the gaps 
and the specific heat shown in Fig.\ref{fig2}.
The presence of
the small gap for the $\gamma$ sheet gives rise to essentially
gapless excitations for temperatures 
$T \gtrsim |{\bf d}_{\gamma}({\bf k})|$ and
this can give rise to the residual density of states observed experimentally.
For temperatures below ${\bf d}_{\gamma}$, this gap gives rise
to the low temperature exponential decay of $C_{es}/T$ to zero.
Note that the density of states is split approximately evenly between
the $\gamma$ sheet and the $\{\alpha,\beta\}$ 
sheets. Consequently,  the smaller gap lying in either the $\gamma$ or 
$\{\alpha,\beta\}$ sheets gives good agreement with the magnitude
of the residual density of states seen experimentally. To show how the 
smaller gap manifests itself in 
other properties we have calculated the London
penetration depth and the thermal conductivity in the basal plane (shown
in Fig.~\ref{fig2} ). The London penetration depth is
\begin{equation}
\lambda^{-2}_{\perp}(T)=\frac{4\pi e^2}{c^2}\frac{1}{\Omega}\sum_{l,{\bf k}}
v_{\perp,l,{\bf k}}^2\left[\frac{\partial f(\epsilon_{l,{\bf k}})}{\partial
\epsilon_{l,{\bf k}}}-\frac{\partial f(E_{l,{\bf k}})}
{\partial E_{l,{\bf k}}} \right].
\end{equation}
which results from a simple extension of the standard BCS expression to 
include many bands.
The thermal conductivity in the single band case is derived in Ref.\cite{arf88}
and the suitable generalization to include many bands is
\begin{equation}
\kappa_{\perp}(T)=-2\sum_{l,{\bf k}}\frac{E_{l,{\bf k}}^2}{T}v_{\perp,l,{\bf k}}^2
\frac{\partial f(E_{l,{\bf k}})}
{\partial E_{l,{\bf k}}} \tau_{l,{\bf k}}
\end{equation}
with
$\tau_{l,{\bf k}}={\tau_{N,l}}
|\epsilon_{l,{\bf k}}|/E_{l,{\bf k}}$
where $\tau_N$ is the normal state 
relaxation time.
This form is valid within the Born approximation. It has been assumed
that there is no interband scattering and  
that $\tau_{N,l}=\tau_N$.
Note that $\tau_{l,{\bf k}}$ does not have the same form as that for 
a conventional isotropic superconductor due to the 
odd-parity coherence factors
 \cite{cof85,pet86}. In calculating these properties
it has been assumed that the density of states corresponds
 to that
of a clean system. However it may be the case 
that while the large gap 
will remain intact in the presence of impurities the smaller 
gap may be rendered gapless (though there will still be a coherent 
pairing amplitude on this Fermi surface sheet \cite{sig91}). 

We have considered a  model in which 
all the Fermi surface sheets favor the same superconducting symmetry.
This model has two order parameters of the same
symmetry (one ($\psi_1$) for the ${\gamma}$ and one ($\psi_2$)
for the $\{\alpha,\beta\}$ sheets) and can in principle have
a second transition from a state in which 
$(\psi_1,\psi_2)=e^{i\theta}(|\epsilon_1|,\pm|\epsilon_2|)$ to a state in 
which time
reversal symmetry is broken: $(\psi_1,\psi_2)=e^{i \theta}(|\epsilon_1|,e^{i
 \phi} |\epsilon_2|)$ where $\phi \ne {0,\pi}$.
An examination of the Ginzburg Landau
coefficients found by a weak coupling analysis shows that the broken time
reversal symmetry phase 
does not occur in this model. However, as was considered by
Leggett for the two band conventional superconductor \cite{leg66}
and more recently by Wu and Griffin in bilayer high $T_c$ superconductors
\cite{wu95},
there 
will exist a collective excitation corresponding to fluctuations into the
broken time reversal symmetry phase 
(fluctuations of the relative phase of $\psi_1$ and $\psi_2$). 
If all orbitals favor the same pairing 
symmetry then 
such a mode may appear below the single particle threshold. 
This mode is in addition to those that were predicted to
exist due to the odd-parity symmetry in the
presence of  weak spin-orbit coupling \cite{ric95}.  
It is also  possible that 
due to the different symmetry properties of the 
the ${xy}$ and the $\{{xz},{yz}\}$ Wannier functions
the $\gamma$ and $\{\alpha,\beta\}$ sheets 
may favor different superconducting symmetries.
In this case a second 
superconducting transition (as opposed to the
crossover behavior shown in Fig.~\ref{fig2}) is likely to occur 
due to the smallness
of the pair scattering amplitude between these two classes of sheets. A   
low temperature broken time reversal symmetry phase
is possible within this scheme \cite{sig91}.

In conclusion, we have presented a model for the superconducting
transition in Sr$_2$RuO$_4$ in which the superconductivity in the 
bands with Ru $d_{xy}$ orbital character   
and the bands with Ru $\{d_{xz},d_{yz}\}$ orbital character  
is weakly coupled. 
This model attributes the large observed residual density of states
to thermal excitations across a secondary gap that is smaller than the
primary gap driving the superconducting transition. This secondary
gap should reveal itself in a wide variety of low temperature 
experiments on sufficiently clean samples. Also, within this model 
a second superconducting transition is possible. Experiments at 
very low temperatures are desirable to examine these possibilities.

We acknowledge the financial support of the Swiss Nationalfonds.
In particular, M.S. is supported by a PROFIL-Fellowship and D.F.A.
by the Zentrum for Theoretische Physik.
D.F.A. also acknowledges financial support from the Natural Sciences
and Engineering Research Council of Canada. 
We also wish to thank Y. Maeno, S. Nishizaki, Y. Kitaoka, and
K. Ishida for useful discussions.

\eject
\vglue 1 cm

\begin{references}
\bibitem{exp} This estimate is $\langle z^2 \rangle^{1/2}$ for a
$4d_{xy}$ orbital with an effective charge $\tilde{Z}=8$.
\end{references}
\end{document}